\documentstyle[11pt,newpasp,twoside,epsf]{article}
\def\spose#1{\hbox to 0pt{#1\hss}}
\def\lta{\mathrel{\spose{\lower 3pt\hbox{$\mathchar"218$}}
     \raise 2.0pt\hbox{$\mathchar"13C$}}}
\def\gta{\mathrel{\spose{\lower 3pt\hbox{$\mathchar"218$}}
     \raise 2.0pt\hbox{$\mathchar"13E$}}}
\def\etal{{\it et al.\ }}

\def\emphasize#1{{\sl#1\/}}
\def\arg#1{{\it#1\/}}
\let\prog=\arg

\def\edcomment#1{\iffalse\marginpar{\raggedright\sl#1\/}\else\relax\fi}
\reversemarginpar

\begin{document}
\title{Young Star Clusters: Metallicities, Ages, and Masses}
 \author{Uta Fritze -- v. Alvensleben}
\affil{Universit\"atssternwarte G\"ottingen, Geismarlandstr. 11, D -- 37083 G\"ottingen, Germany}

\begin{abstract}
Observations of Young Star Cluster ({\bf YSC}) systems in interacting
galaxies are reviewed with particular emphasis on their
Luminosity Functions ({\bf LF}) and colour distributions. A few spectroscopic
abundance measurements are available. They will be compared to YSC
abundance predictions from spiral galaxy models. Evolutionary
synthesis models allow to derive ages for individual YSCs on the basis
of their broad band colours. With individual YSC ages models
predict the future colour and luminosity evolution of the YSC systems
that will be compared -- after a Hubble time -- to observations of old
Globular Cluster ({\bf GC}) systems. Using model M/L ratios as a function of age,
YSC masses can be estimated. Age spread effects in young systems can
cause the shape of the LF to substantially differ from the shape of
the underlying mass function. Major sources of uncertainty are the
metallicity, dust reddening, and observational colour uncertainties.  
\end{abstract}


\section{Why do we want to know GC and YSC masses?}

GCs are conventionally believed to be (among) the oldest objects in the Universe, 
dating back to the times of galaxy formation. They are used to constrain 
the age of the Universe. LFs of GC systems are believed to be universal 
enough for determinations of distances to $> 20$ Mpc and of the Hubble 
constant. On the other hand, ``present-day GCs are the hardiest survivors of 
a larger original population'' (Harris 1991). Hence, their observed LF is not 
only their LF at formation shifted by stellar evolutionary fading, but might be 
additionally modified by cluster destruction processes. Dynamical modelling 
of cluster systems in the Galactic potential shows that destruction processes 
and timescales strongly depend on cluster masses. While destruction by dynamical 
friction is more efficient for high mass clusters, tidal shocking and 
evaporation preferentially destroy low mass clusters. 

Bright YSCs are observed in large numbers in interacting galaxies and 
merger remnants and a burning question with far-reaching implications is 
if these YSCs are young GCs. Masses derived for these YSCs are much higher 
than those of open clusters in the Milky Way. Effective radii of YSCs typically 
are a few pc similar to GC radii. The 
mass function ({\bf MF}) of YSC systems in comparison with the MFs of 
molecular clouds or molecular cloud cores tells us about star and cluster 
formation processes. The MF of YSCs, as first pointed out by 
Meurer (1995), may differ in shape from 
their LF since M/L varies rapidly at young ages and the age spread 
within a YSC system is not much smaller than its age. The LF of open clusters in the Milky Way, the MFs of 
molecular clouds and molecular cloud cores, the observed LFs of YSCs, 
e.g. in 
NGC 4038/39, NGC 7252, NGC 3256, and of giant HII regions 
all are power laws with slopes in the range  
$\alpha \sim -1.5 \, . \, . \, . \, -1.8$ (cf. Solomon \etal 1987, Lada \etal 1991, 
Kennicutt 1989, and reviews by Harris \& Pudritz 1994, Elmegreen \& Efremov 1997). 
Yet the LF of {\bf old} GCs is Gaussian with typical parameters 
${\rm \langle M_V \rangle \sim -7.3 \, mag, ~\sigma (M_V) \sim 1.3 \, mag}$, 
their MF is log-normal with typically ${\rm \langle Log(M/M_{\odot}) \rangle 
\sim 5.5, ~ \sigma \sim 0.5}$ (e.g. Ashman \etal 1995).

Hence the question as to the MF of YSCs has profound implications. 
If the MF of YSCs were a power law like their LF and if YSC systems are 
to evolve into something similar to old GC systems, dynamical 
destruction processes would have to transform the power law MF into a 
log-normal MF over a Hubble time. If, on the other hand, the MF of YSCs were log-normal 
(and their LF distorted to a power law by age spread effects), then the 
star/cluster formation process would have to transform the power law MF of the 
molecular clouds into the log-normal MF of YSCs. Or, else, might already the MF of 
molecular clouds/cloud cores in violently star forming mergers (where to my 
knowledge it has not yet been observed) be different 
from what it is in quietly star forming `normal galaxies' (where it is observed)? 

\section{Evolutionary synthesis of SSPs}
Star clusters are simple stellar populations ({\bf SSP}s), formed in one 
short burst of star formation with one metallicity Z. 
With evolutionary synthesis models, the time evolution of SSPs of 
various metallicities ${\rm Z_i}$ is studied in terms of luminosities ${\rm L_{\lambda}}$, 
colours, spectrum, absorption features, stellar mass loss, and hence M/L by 
many groups (e.g. Bruzual \& Charlot 1993, Worthey 1994, Bressan \etal 1994, 
F.-v.A. \& Burkert 1995, Leitherer \etal 1999, Kurth \etal 
1999). Basic parameters of this approach are the IMF and the set of stellar 
evolutionary tracks used (e.g. from the Padova or Geneva groups). 
All models agree that the changes in luminosities and colours are 
rapid in early evolutionary stages and become slower with increasing age. The 
colour evolution depends significantly on metallicity, already in 
very early stages, and the fading also depends on metallicity, in particular 
during the first Gyr (cf. F.-v.A. \& Burkert 1995, Kurth \etal 1999).

\section{YSC observations vs. SSP models}
\subsection{Procedure}
If the metallicity of YSCs is known, individual ages can be obtained on 
the basis of their observed UBVI colours. Unfortunately, metallicity information from 
spectroscopy is only available yet 
for a handful of YSCs in NGC 7252 (Schweizer \& Seitzer 1993, 1998) and NGC 1275 (Brodie \etal 1998). 
In all other cases we have to go back to some educated guess. In gas rich galaxy mergers, 
YSCs form out of the gas from 
the progenitor (spiral) galaxies, the abundance of which gives a lower limit to 
YSC abundances. E.g., for the YSCs in NGC 7252, which is a merger of two luminous 
Sc type spirals, we had predicted ${\rm Z_{YSC} \ga Z_{ISM}^{Sc} \sim \onehalf 
Z_{\odot}}$ from the ISM abundance evolution in our 1-zone spiral models (F.-v.A. \& Gerhard 1994). 
Spectroscopy of the brightest YSCs yielded ${\rm Z_{YSC} \sim Z_{\odot}}$ with some tentative 
indication of self-enrichment during the burst from the comparison of Mgb and Fe lines 
(cf. F.-v.A. \& Burkert 1995). 
Distances are known for all YSC systems. Hence absolute luminosities ${\rm L_V}$ can be 
combined with ${\rm M/L_V -}$ values from SSP models of appropriate age and metallicity to 
derive the masses of YSCs. 

\subsection{Sources of uncertainty}
Sources of uncertainty for these mass estimates come both from observations and models. 
Observational errors on luminosities and colours might be inhomogeneous and probably 
not independent of each other, dust extinction is not known for individual YSCs, the dust 
distribution in some systems is inhomogeneous, the metallicity is, at most, known for 
a small subsample of YSCs and might show an intrinsic scatter, and, finally, the 
completeness limit need not be homogeneous over the region of YSC observations. 
Intrinsic model uncertainties are estimated to be $\la 0.1$ mag in 
(optical) luminosities and colours. Differences between models from various authors 
are mainly due to differences in the stellar input physics. While serious colour 
discrepancies at very young ages $\sim 10$ Myr are seen, e.g. comparing models from 
Bruzual \& Charlot with those of Leitherer \etal, probably due to the inclusion/non-inclusion 
of emission lines, {\bf fairly good agreement is reached among models from various groups 
for the same metallicity and IMF at all 
ages $\ga 60 - 100$ Myr}. E.g., ${\rm \Delta(V-I) \mid_{12\,Gyr} \la 0.1}$ mag and 
${\rm \Delta(M/L_V) \mid_{12\,Gyr} \la 10 \%}$. 

${\rm M/L_{\lambda}}$, however, depends on wavelength $\lambda$, metallicity Z, and IMF, i.e. on 
its slope and lower mass limit. 
In Tab.1, I briefly sketch out these dependencies as obtained from our models 
using Padova stellar evolutionary tracks for stars in the mass range ${\rm 0.1 - 60~ M_{\odot}}$ 
(cf. Kurth \etal 1999) and 
two different IMFs (Salpeter vs. Scalo 1986). 

In general and as reported by others before, model ${\rm M/L_V}$ are about twice 
as large as are ${\rm M/L_V}-$ values derived from observations of old GCs, even when a 
Scalo IMF is assumed. For a Salpeter IMF the discrepancy is higher by another factor of two. 
Observational ${\rm M/L-}$ values are obtained by 
measuring the central velocity dispersion $\sigma_0$, central surface brightness ${\rm I_0}$, 
and half light radius ${\rm r_{1/2}}$, and assuming isotropic orbits, no radial gradients 
in ${\rm M/L}$, and no DM halos around GCs. Then 
\begin{center} ${\rm M/L \sim \frac{\sigma_0^2}{I_0 \, r_{1/2}}}$ \end{center}
and typical values quoted for Galactic GCs are ${\rm M/L_V \sim 2}$. 
There are two effects that might invalidate the assumptions going into this derivation. 
First of all, mass segregation in GCs (cf. Meylan {\sl this conf.}) will 
result in an ${\rm M/L}$ increasing with 
radius, as e.g. observed by C\^{o}t\'e \etal 1995 for NGC 3201. Second, low mass stars with 
their high ${\rm M/L}-$ values are preferentially 
lost by evaporation (e.g. Gerhard {\sl this conf.}). 
While both processes are undoubtedly at work in GCs, attempts 
to examine quantitatively the validity of the assumptions 
involved seem to still give ambiguous results. Leonard \etal 1992 use 
proper motion 
data and radial velocity measurements for stars in the GC M13 and find that 
$\sim 50 \%$ of the 
mass of M13 is in low mass stars and brown dwarfs. Including this unseen mass 
leads to an increased 
${\rm M/L \sim 4}$, a value well compatible with the models. On the other hand, 
observations of tidal tails on some GCs in the Milky Way potential are interpreted to 
indicate that there is not much DM around those GCs, constraining their 
mass-to-light ratios to 
${\rm M/L \la 2.5}$ (Moore 1996). 

\begin{table}[h]
\vspace{-6pt}
\caption{M/L-values from SSP models at various wavelengths for young, intermediate age, 
and old stellar populations compared for two metallicities and two different IMFs.}
\vspace{0.4truecm}
\begin{tabular}{|l|cccccc|} \hline
	& & & & & & \\ 
   & & & ${\rm Z=10^{-3}}$ & ${\rm 2 \cdot Z_{\odot}}$ & & \\ 
	& & & & & & \\ 
	\hline 
	& & & & & & \\ 
 $10^8$ yr & ${\rm M/L_{U,B}}$ & $\sim$ & 0.1 & 0.1 & & indep. of \\ 
	& ${\rm M/L_{V,R}}$ & $\sim$ & 0.2 & 0.2 & & Z \& IMF \\ 
	& ${\rm M/L_{K}}$ & $\sim$ & 0.3 & 0.5 & & indep. of IMF \\ 
	& & & & & & \\ 
	\hline 
	& & & & & & \\ 
  $10^9$ yr& ${\rm M/L_B \mid _{Salp}}$ & $\sim$ & 0.5 & 0.9 & $\sim$ & ${\rm 2 \times M/L_B \mid _{Scalo}}$ \\ 
  & ${\rm M/L_V \mid _{Salp}}$ & $\sim$ & 0.8 & 1.0 & $\sim$ & ${\rm 2 \times M/L_V \mid _{Scalo}}$ \\ 
  & ${\rm M/L_K \mid _{Salp}}$ & $\sim$ & 1.5 & 0.9 & $\sim$ & ${\rm 1.5 \times M/L_K \mid _{Scalo}}$ \\ 
	& & & & & & \\ 
	\hline 
	& & & & & & \\ 
  12 Gyr & ${\rm M/L_{B,V} \mid _{Salp}}$ & $\sim$ & 8 & 12 & $\sim$ & ${\rm 2 \times M/L_{B,V} \mid _{Scalo}}$ \\ 
  & ${\rm M/L_K \mid _{Salp}}$ & $\sim$ & 5 & 1.5 &  & -- \\ 
  & ${\rm M/L_K \mid _{Scalo}}$ & $\sim$ & 3.5 & 2.7 &  & -- \\ 
	& & & & & & \\ 
	\hline 
  \end{tabular}
\vspace{-6pt}
\end{table}

\section{YSCs in the Antennae: a 1$^{st}$ example}
The Antennae galaxies (= NGC 4038/39) is an interacting pair of two gas rich spirals, 
probably Sc, of 
comparable mass. In the ongoing starburst triggered by the interaction a population 
of bright YSCs is formed, numerous enough to allow for the first time to reasonably 
define a LF. The LF from WFPC1 observations is 
a power law with slope $\alpha \sim -1.8$ (Whitmore \& Schweizer 1995 (WS95)). 
WFPC2 reobservations are going to be presented by Miller ({\sl this conf.}).

Assuming a homogeneous metallicity ${\rm Z \sim 1/2 \cdot Z_{\odot}}$ lack of individual 
cluster spectroscopy and in analogy to the YSC system in NGC 7252, I analysed the WFPC1 
data of WS95 with evolutionary synthesis models for SSPs. 
In a first step, the average dereddened ${\rm (V-I)}$ colour of the 550 YSCs 
is used to derive a mean age of 
the YSC population of $(2 \pm 2) \cdot 10^8$ yr, consistent with Barnes' (1988) dynamical 
model for the interaction between NGC 4038 and 4039 and consistent with Kurth's (1996) 
global starburst age. 
SSP models describe both the fading and the reddening of the YSC population 
as it ages. Assuming a mean age for the YSC population, both the LF and the colour 
distribution of the YSCs are simply shifted towards fainter magnitudes and redder 
colours, respectively, without changing shape.

\subsection{Evolution of the YSC LF}
In a second step, individual ages are derived for all the YSCs on the basis of their 
individual ${\rm (V-I)}$ and ${\rm (U-V)}$ colours. Interestingly, the resulting 
age distribution not only 
shows a peak at very young ages $0 - 4 \cdot 10^8$ yr for the YSCs, but also a contribution from 
$\sim 12$ Gyr old GCs from the parent galaxies (Fig. 1a). 
A small number of apparent interlopers are 
probably due to inhomogenities in the dust distribution. It is improbable that many of the old GCs 
we identify are highly reddened YSCs, since they would have to be exeedingly bright 
intrinsically. The fact that age estimates from ${\rm (U-V)}$ colours, as far as available, agree 
with age estimates from ${\rm (V-I)}$ supports our metallicity assumption and makes us hope that 
the average ${\rm E_{B-V}}$ is correct for most of the clusters. It is clear, however, 
that the individual YSC extinctions 
are a major source of uncertainty in this analysis. 
With individual YSC ages and observed luminosities, SSP models can be used to 
predict the 
time evolution of cluster luminosities. Neglecting any kind of dynamical cluster 
destruction effects, and only using the {\bf young} star clusters identified, 
we obtain the surprising result that {\bf by an age of 12 Gyr}, when age differences 
among individual YSCs will be negligible, {\bf their LF will 
have evolved from the presently observed power law 
into a fairly normal Gaussian GC LF} with parameters ${\rm M_V = -6.9}$ mag 
and ${\rm \sigma (M_V) = 1.3}$ mag (Fig. 1b). 
The turn-over then is $\ga 1$ mag brighter than the completeness limit which 
evolves to ${\rm M_V = -5.7}$ mag, and fainter by $\sim 0.4$ mag than for 
typical GC systems. This is a consequence of the enhanced metallicity 
of the YSC population with respect to that of old GC systems, and in agreement 
both with observations of old GC systems with a range of metallicities 
(Ashman \etal 1995) and with our SSP model predictions. 
With this surprising result we confirm and quantify Meurer's conjecture that 
{\bf over a Hubble time, age spread effects can transform an observed power law LF of 
YSCs into the Gaussian LF of old GCs} (cf. F.-v.A. 1998 for details). 

The bright end of the LF defined by the old GCs from the parent spirals is well 
described by a Gaussian LF with parameters 
${\rm \langle M_V \rangle = -7.3 }$ mag and ${\rm \sigma (M_V) = 1.2 \, mag,}$ 
normalised to the total number of GCs in the Milky Way and Andromeda galaxies together (Fig. 2a). 
Hence, if the two interacting spirals NGC 4038 and 4039 had a similar number of 
GCs as those galaxies, and if the bulk of the YSC population really are young GCs, 
Zepf \& Ashman's (1993) requirement, that the number of 
secondary GCs formed in mergers should be comparable to the number of primary GCs in 
the two progenitor spirals, would be fulfilled. This requirement is necessary if 
the higher specific GC frequency 
in ellipticals -- as compared to spirals -- is to be compatible with a spiral-spiral merger 
origin for those ellipticals.

\begin{figure}[t]
\plottwo{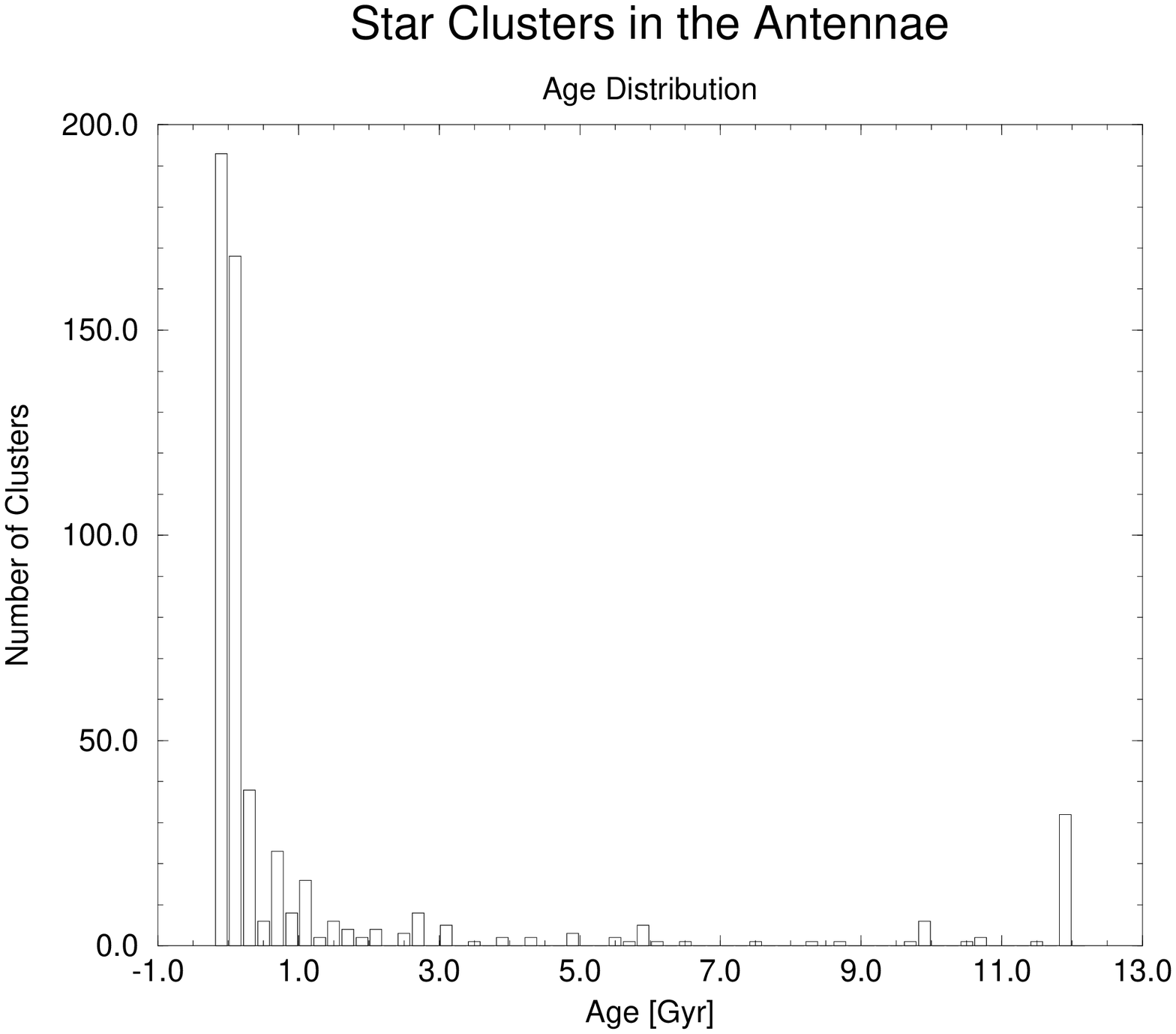}{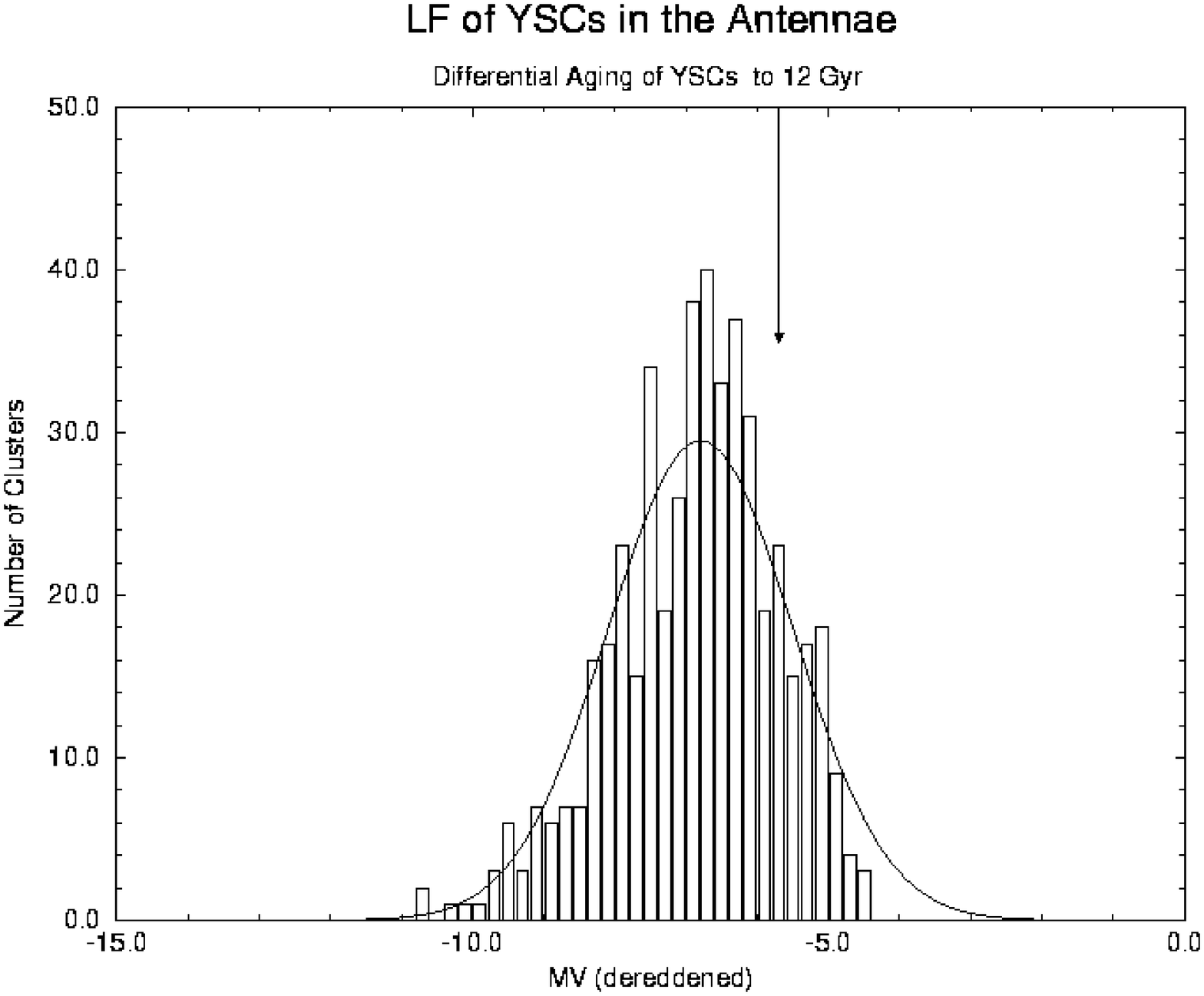}
\caption{(a) Age distribution of {\bf all} clusters, (b) LF evolved to 12 Gyr by differential fading for 
{\bf young} clusters in the Antennae. The vertical arrow indicates the completeness limit.}
\end{figure}
\begin{figure}[t]
\plottwo{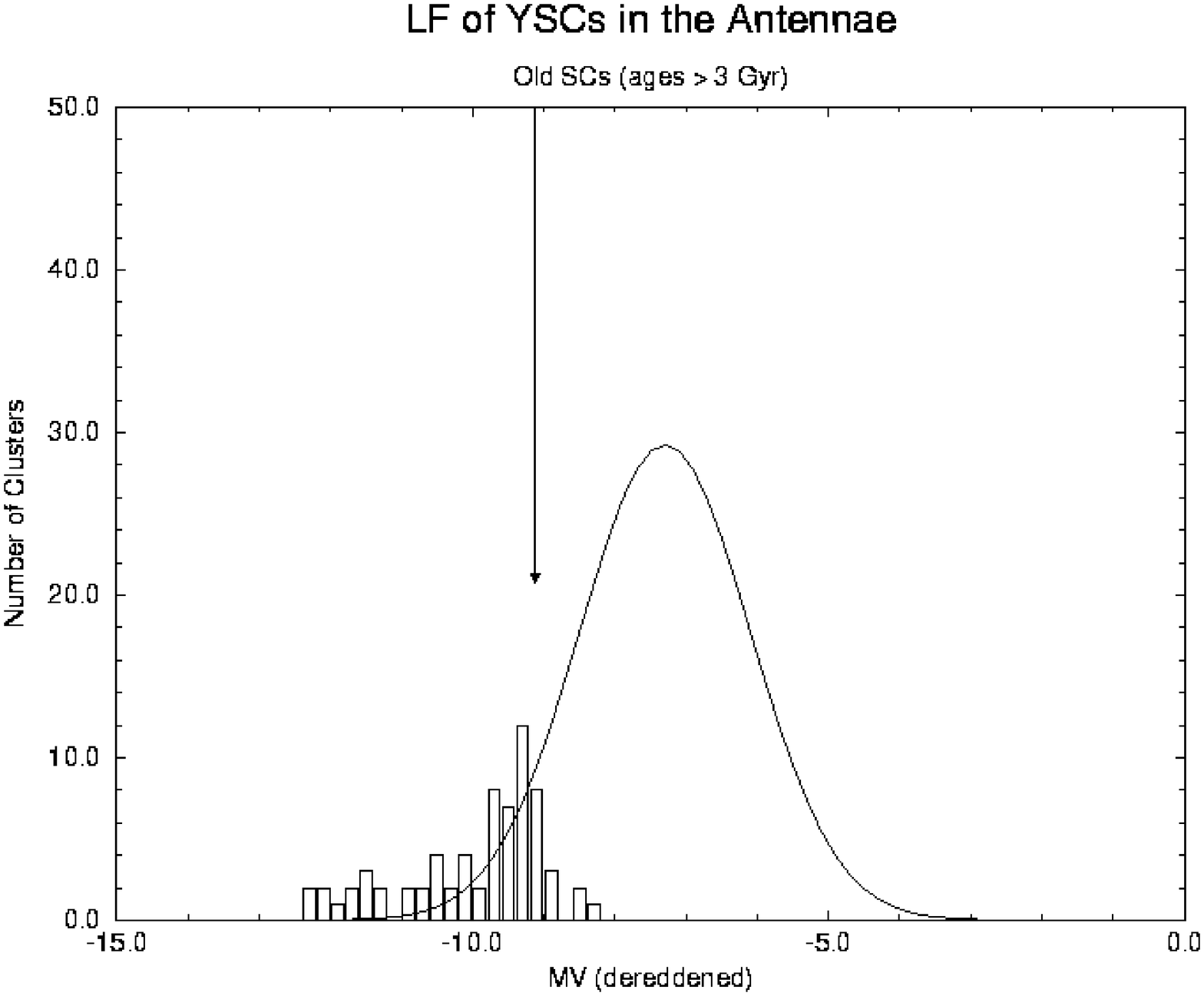}{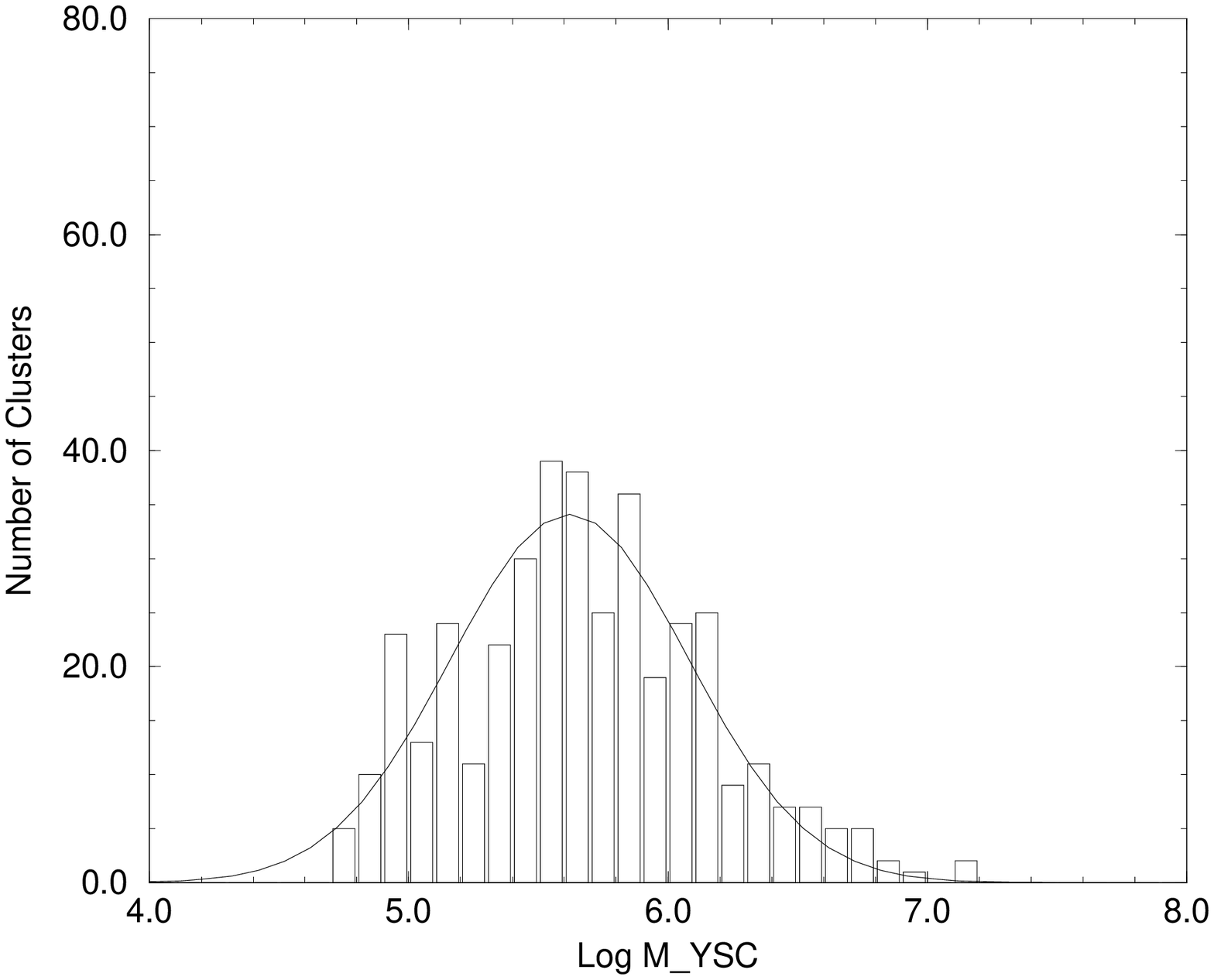}
\caption{(a) LF of old GCs with Gaussian for GCs in (Milky Way + M31).  
(b) MF of the young clusters.}
\end{figure}

\subsection{MF of YSCs}
To investigate whether the LFs and MFs of old GC systems are determined by 
the cluster formation process or whether they are the result of secular 
dynamical destruction processes, we derive 
the MF of the very young star cluster system in the Antennae. 
We restrict ourselves to YSCs brighter than the completeness limit, use 
individual YSC ages and luminosities, and combine them with model M/L for the 
respective YSC ages to determine their individual masses. {\bf Our tentative 
conclusion is that for all the 393 YSCs brighter than the completeness limit 
and with ${\rm (V-I)}$ colours available, the MF is compatible with a Gaussian MF} 
with parameters 
${\rm \langle Log(M_{YSC}/M_{\odot}) \rangle = 5.6,~~\sigma \sim 0.46}$ 
${\rm \longleftrightarrow ~~ \langle Mass(YSC) \rangle \sim 4 \cdot 10^5~M_{\odot}}$ (Fig. 2b). 
If we include 
YSCs fainter than the completeness limit, neither the shape nor the parameters of 
the MF are changed (cf. F.-v.A. 1999 for details).

The uncertainty in the YSC metallicity leads to age uncertainties, which, 
in turn, lead to uncertainties in model M/L values of the order of 10 \% in 
these early stages. Inhomogenities in the dust distribution do not seem to be very 
important for the brightest YSCs for which ages from ${\rm (U-V)}$ agree with ages 
from ${\rm (V-I)}$ colours. We do not know, however, how important they are for 
those YSCs that are not detected in U. Hence, on the basis of ${\rm (V-I)}$ alone, 
we cannot quantify in how far an inhomogeneous internal reddening might affect 
our results. 
For this reason and since our analysis is based on WFPC1 data, we caution that 
our conclusions concerning the evolution of the YSC LF and their MF can only be 
preliminary. As soon as the reduced WFPC2 data presented by Miller will become 
available to us, we shall repeat our analysis and supplement it with simulations 
to estimate the effects of differential observational uncertainties. 

\subsection{Implications}
If, however, our preliminary result became confirmed, this would imply that the 
log-normal MF of old GCs is produced by the cluster formation process rather 
than by secular dynamical evolution of the cluster system. In this context. 
it seems very interesting to obtain observational information about the molecular 
cloud mass spectrum in massive interacting galaxies. Jog \& Solomon (1992) conjecture that 
the strongly enhanced ambient pressure in massive gas rich mergers might affect the 
molecular cloud structure. In Ultraluminous Infrared Galaxies --  which all are 
mergers with strong starbursts -- the fraction of gas at very high densities of 
${\rm n \sim 10^4 ~and~ 10^5 \, cm^{-3}}$, as traced by HCN and CS lines, 
with respect to gas at ${\rm n \sim 500 \, cm^{-3}}$, as traced by CO, is indeed 
observed to be higher by 1 -- 2 orders of magnitude (e.g. Solomon \etal 1992). 

\subsection{Discussion}
Even with deeper WFPC2 data, however, the MF of the YSCs in the Antennae does 
not yet seem to be unambiguously settled. Zhang \& Fall (1999) use 
reddening-free colour indices Q$_1$, Q$_2$, solar metallicity Bruzual \& Charlot 
models, consider incompleteness and stellar contamination, and find power law 
MFs with slopes $\alpha \sim -2$ for YSCs in the two age intervals 2.5 -- 6.3 Myr 
and 25 -- 160 Myr where Q$_1$ and Q$_2$ give unambiguous results. 
In an independent analysis of the same data, Miller ({\sl this conf.}) 
finds a significant flattening in the observed LF at ${\rm M_V \sim -10.7}$ mag. Intriguingly, 
this value is exactly the turnover luminosity implied by our MF at the mean YSC age in 
the absence of an age spread. Miller's method of reconstruction, however, leads 
to a power law YSC MF. Clearly, more work both from the modelling and observational 
sides is needed before we really understand the YSC MF. 

\subsection{Other systems}
For the YSCs in NGC 7252 and NGC 3921, Miller \& Fall (1997) and Miller ({\sl this conf.}) 
prefer power law MFs. 
Distances to those systems, however, are larger by factors 3 and 4, respectively, 
pushing the completeness limit to higher luminosities, even in WFPC2 data. 
Moreover, the YSCs in these 
dynamically old merger remnants have higher mean ages (650 -- 750 Myr for NGC 7252, 
250 -- 750 Myr for NGC 3921, as compared to 200 Myr for the Antennae), and hence, 
are intrinsically fainter already 
by 0.8 -- 1 mag, on average. 
If we assume that the MFs for the YSCs in NGC 7252 and NGC 3921 were identical to the 
one we derived for the YSCs in the Antennae, we obtain turn-over luminosities 
${\rm \langle M_V \rangle = -10 \dots -9.5}$ mag for NGC 7252 and 
${\rm \langle M_V \rangle = -9.5 \dots -9}$ mag for NGC 3921 at their respective mean 
YSC ages. In both cases, these turn-over luminosities are close to 
the completeness limits. The difficulty to disentangle the old GC and 
the YSC populations also increases with increasing YSC age. We therefore doubt 
that it is possible to track the MF beyond a possible turnover in NGC 7252 and 3921. 
In any case, the YSCs in NGC 7252 and 3921 have survived $\gg 10$ crossing times and 
from this fact alone are young GCs rather that open ones, as also indicated by their 
combination of small effective radii and high luminosities (cf. Miller \etal 1997, 
Schweizer \etal 1996). 

\subsection{Dynamical Evolution}
Stellar mass loss is included in our models for 
the evolution of M/L. Over 12 Gyr, the masses of YSCs will decrease by ${\rm \sim 10 ~ and ~ 15 \%}$ 
for a Salpeter and Scalo-IMF, respectively. About half of the entire stellar 
mass loss occurs during the ${\rm 1^{st}}$ Gyr. 
External dynamical effects from an interaction of the clusters with the potential 
of the interacting galaxy system seem extremely difficult to model. Comparison of YSC 
LFs and MFs in an age sequence of interacting galaxies, mergers, merger remnants, and 
dynamically young ellipticals 
will allow to ``see these processes at work''. 

In the Milky Way potential, Vesperini (1998) has shown that an assumed initially 
log-normal GC MF is conserved in shape and parameters during self-similar evolution over a Hubble 
time despite the destruction of $\sim 50 \%$ of the 
cluster population. If, on the 
other hand, he starts with a power law MF, severe fine-tuning is required for his 
model parameters to secularly transform it into the observed log-normal MF 
of old GC systems. 

It is clearly important to analyse more YSC systems in order to see if (and in how far) 
their MFs are universal or might depend on environment. Old GC systems have their 
turn-over around ${\rm \langle M_V \rangle \sim -7.2}$ mag. With 10m 
telescopes they are accessible to more than Virgo cluster distances. MOS -- e.g. 
with FORS on the VLT -- in combination with HST imaging will 
allow to determine cluster abundances and hence to more precisely age-date them, and may 
even provide kinematic information for independent mass estimates. 

An open question seems to me if the (globular) cluster formation process in 
the high metallicity environment of interacting spirals today is the same or not as it was 
in the Early Universe when the radiation field was stronger and the metallicity lower. 

\acknowledgements
I wish to thank the organisers and in particular Ariane Lan\c{c}on for a stimulating workshop 
in a very pleasant and warm atmosphere in cold and rainy Strasbourg.


\begin{references}

\reference Ashman, K. M., Conti, A., Zepf, S. E., 1995, AJ 110, 1164
\reference Barnes, J. E., 1988, ApJ 331, 699
\reference Bressan, A., Chiosi, C., Fagotto, F., 1994, ApJS 94, 63
\reference Brodie, J. P., Schr\"der, L. L., Huchra, J. P., \etal 1998, AJ 116, 691
\reference Bruzual, G. A., Charlot, S., 1993, ApJ 405, 538
\reference C\^{o}t\'e, P., Welch, D. L., Fischer, P., Gebhardt, K., 1995, ApJ 454, 788
\reference Elmegreen, B. G., Efremov, Y. N., 1997, ApJ 480, 235
\reference Fritze -- v. Alvensleben, U., 1998, A\&A 336, 83
\reference Fritze -- v. Alvensleben, U., 1999, A\&A 342, L25
\reference Fritze -- v. Alvensleben, U., Burkert, A., 1995, A\&A 300, 58
\reference Fritze -- v. Alvensleben, U., Gerhard, O. E., 1994, A\&A 285, 751 + 775
\reference Harris, W. E., 1991, ARA\&A 29, 543
\reference Harris, W.E., Pudritz, R. E., 1994, ApJ 429, 177
\reference Jog, C. J., Solomon, P. M., 1992, ApJ 387, 152
\reference Kennicutt, R. C., 1989, ApJ 344, 685
\reference Kurth, O., 1996, Diploma Thesis, Univ. G\"ottingen
\reference Kurth, O., Fritze - v. Alvensleben, U., Fricke, K. J., 1999, A\&AS 138, 19
\reference Lada, E., Bally, J., Stark, A. A., 1991, ApJ 368,432
\reference Leitherer, C., \etal 1999, ApJS 123, 3
\reference Leonard, P. J. T., Richer, H. B., Fahlman, G. G., 1992, AJ 104, 2104
\reference Meurer, G. R., 1995, Nat 375, 742
\reference Miller, B. W., Fall, S. M., 1997, AAS 119, \#115.04
\reference Moore, B., 1996, ApJ 461, L13
\reference Scalo, J. M., 1986, Fundam. Cosm Phys. 11, 1
\reference Schweizer, F., Seitzer, P., 1993, ApJ 417, L29
\reference Schweizer, F., Seitzer, P., 1998, AJ 116, 2206
\reference Schweizer, F., Miller, B. W., Whitmore, B. C., Fall, S. M., 1996, AJ 112, 1839
\reference Solomon, P. M., Rivolo, A. R., Barrett, J., Yahil, A., 1987, ApJ 319, 730
\reference Solomon, P. M., Downes, D., Radford, S. J. E., 1992, ApJ 387, L55
\reference Vesperini, E., 1998, MN 299, 1019
\reference Whitmore, B.C., Schweizer, F., 1995, AJ 109, 960 (WS95)
\reference Worthey, G., 1994, ApJS 95, 107
\reference Zepf, S. E., Ashman, K. M., 1993, MN 264, 611
\reference Zhang, Q., Fall, S. M., 1999, ApJ 527, L81

\end{references}
\end{document}